\documentclass[pra,twocolumn,superscriptaddress,noshowpacs]{revtex4-1}
\usepackage{amssymb}
\usepackage{amsmath}
\usepackage{amsfonts}
\usepackage{mathtools}
\usepackage[mathlines]{lineno}
\usepackage{bm}
\usepackage{dcolumn}
\usepackage{physics}
\usepackage{graphicx}
\usepackage[dvipsnames]{xcolor}
\usepackage{color}
\usepackage{xcolor}
\usepackage{calc} 
\usepackage{graphics}
\usepackage{epsfig}
\usepackage{epstopdf}
\usepackage{float}
\usepackage{natbib}
\usepackage[colorlinks, linkcolor=blue, urlcolor=blue, anchorcolor=blue, citecolor=blue]{hyperref}
\usepackage{color}
\usepackage{diagbox}
\usepackage{makecell}
\usepackage{multirow}
\usepackage{caption}
\begin{document}
	
	\title{Non-Abelian Aharonov-Bohm Caging in Photonic Lattices}
	
	\author{Sheng Li}
	\affiliation{School of Physics, Huazhong University of Science and Technology, Wuhan, 430074, China}
	
	\author{Zheng-Yuan Xue}
	\affiliation{Guangdong Provincial Key Laboratory of Quantum Engineering and Quantum Materials, GPETR Center for Quantum Precision Measurement, and School of Physics and Telecommunication Engineering, South China Normal University, Guangzhou 510006, China}
	
	\author{Ming Gong}
	\affiliation{CAS Key Laboratory of Quantum Information, University of Science and Technology of China, Hefei 230026, China}
	
	\author{Yong Hu}
	\email{huyong@mail.hust.edu.cn}
	\affiliation{School of Physics, Huazhong University of Science and Technology, Wuhan, 430074, China}
	
	\begin{abstract}
    Aharonov-Bohm (AB) caging is the localization effect in translational-invariant lattices due to destructive interference induced by penetrated magnetic fields. While current research focuses mainly on the case of Abelian AB caging, here we go beyond and develop the non-Abelian AB caging concept by considering the particle localization in a 1D multi-component rhombic lattice with non-Abelian background gauge field. In contrast to its Abelian counterpart, the non-Abelian AB cage depends on both the form of the nilpotent interference matrix and the initial state of the lattice. This phenomena is the consequence of the non-Abelian nature of the gauge potential and thus has no Abelian analog. We further propose a circuit quantum electrodynamics realization of the proposed physics, in which the required non-Abelian gauge field can be synthesized by the parametric conversion method, and the non-Abelian AB caging can be unambiguously demonstrated through the pumping and the steady-state measurements of only a few sites on the lattice. Requiring only currently available technique, our proposal can be readily tested in experiment and may pave a new route towards the investigation of exotic photonic quantum fluids.
	\end{abstract}

\maketitle
\section{Introduction}
\label{Sec Intro}
Synthetic gauge fields in artificial atomic and photonic systems has been studied extensively in the past decades \cite{GoldmanGauge2014RPP,GoldmanGauge2018}.
The motivation is to investigate exotic topological physics \cite{BernevigTIBook2013} in a well-controlled quantum simulator \cite{TopoAtom2019RMP,TopoPhoton2019RMP}. Inspired by the rapid progress in this field, research attention has been recently devoted to the realization of Aharonov-Bohm (AB) caging, where the interplay of the external magnetic field and the lattice geometry leads to completely flat bands (FB) in the 1D rhombic lattice \cite{Longhi:14,PhysRevLett.85.3906,PhysRevLett.88.227005,NPRhombic2009} and 2D dice lattice \cite{VidalABCPRL1998} and consequently exotic localization on the perfectly periodic lattices. Being distinct from the localization due to disorder, this effect is interpreted by the destructive interference induced by the Peierels phase of the penetrated magnetic flux. The interaction mechanism then becomes dominant on the dispersionless band, making the AB caging lattices ideal platforms of exploring strongly correlated physics. Following the experimental realization of AB caging early in solid state systems \cite{2DABCPRL1999,2DABCPRL2001} and recently in photonic systems \cite{MukherjeeABCRealization2018PRL,KremerABC2018}, a variety of related theoretical works has been presented, including the effects of nonlinearity and disorder \cite{PhysRevB.64.155306,ABCPair2018PRB,ABCBoseHubbard2018,MaluckovNLABC2019PRA,Di2018Nonlinear,Kuno_2020}, topological pumping and edge states \cite{ABCAtom2019PRA,ABCEdge2019PRA,ABCPumping2019CP}, FB laser \cite{Longhi:19}, and influence of non-Hermicity \cite{ABCFlatBand2019PRA,ABCFlatBand2019PRA2}.

Meanwhile,  subtlety still exists in the sense that, discussions up to now are mainly based on the assumption that an Abelian background gauge potential is imposed. Few works have investigated models in which Rashba SOC can lead to flat band localization in certain lattice geometries \cite{RashbaCage2004PRL,RashbaCage2005PRB}. On the other hand, non-Abelian gauge field \cite{KogutNAGRMP1983} has already manifested its essence in condensed matter physics and quantum optics, partially by the role of spin-orbital coupling (SOC) \cite{SpielmanSOC2013Nature} in the physics of topological quantum matters \cite{BernevigTIBook2013}. In addition, synthetic SOC has been experimentally implemented in various artificial system during the past few years, ranging from ultra-cold atoms \cite{PanJWSOCScience2016} and exciton-polariton microcavities \cite{AmoSOCPRX2015} to coupled pendula chains \cite{SalernoSOC2017NJP}. These exciting advances thus raise the curious questions that whether the AB caging concept can be extended to the realm of non-Abelian gauge, and whether the non-Abelian nature of the gauge field would bring any new physics that has no Abelian correspondence.

In this manuscript, we propose the concept non-Abelian AB caging (i. e. AB caging in the presence of non-Abelian background gauge field) in a 1D rhombic lattice where the lattice sites contain multiple (pseudo)spin components and the background gauge potential becomes matrix-valued. In particular, the non-Abelian AB caging is defined by the emergence of nilpotent interference matrix, which is the matrix generalization of the destructive interference condition in the Abelian AB caging case. Detailed analysis further implies that particle localization in this situation is drastically different from its Abelian counterpart: The non-Abelian feature of the background gauge field results in exotic spatial configuration of non-Abelian AB cage, which is sensitive to both the nilpotent power of the interference matrix and the initial state of the lattice.

In addition, we consider the circuit quantum electrodynamics (QED) system \cite{LiuYXReview2017PR} as a promising candidate of implementing the proposed physics. While the Abelian AB caging has been realized in a variety of physical systems \cite{MukherjeeABCRealization2018PRL,KremerABC2018,PhysRevLett.83.5102,PhysRevLett.85.3906,PhysRevLett.86.5104}, our proposal takes the advantages of
flexibility and tunability of superconducting quantum circuit, which allow the future incorporation of photon-photon interaction and disorder in a time- and site- resolved manner \cite{KochReview1,KochReview2}. The required link variables can be synthesized by the parametric frequency conversion (PFC) approach \cite{NISTParametricConversionNP2011,GrossPFCEPJQT2016,RoushanChiral2017NP}, which is feasible with current technology and can lead to \textit{in situ} tunability of the synthesized non-Abelian gauge potential. Our numerical simulations pinpoint that localization-in-continuum dynamics can be observed through the steady-state photon number (SSPN) detection of only few sites on the lattice, which can serve as unambiguous evidence of the non-Abelian AB caging.

\section{non-Abelian AB Caging: concept and examples}
\label{Sec NAABC}	
\subsection{Definition of non-Abelian AB caging}
\begin{figure}[tbhp!]
\begin{center}
\includegraphics[width=0.49\textwidth]{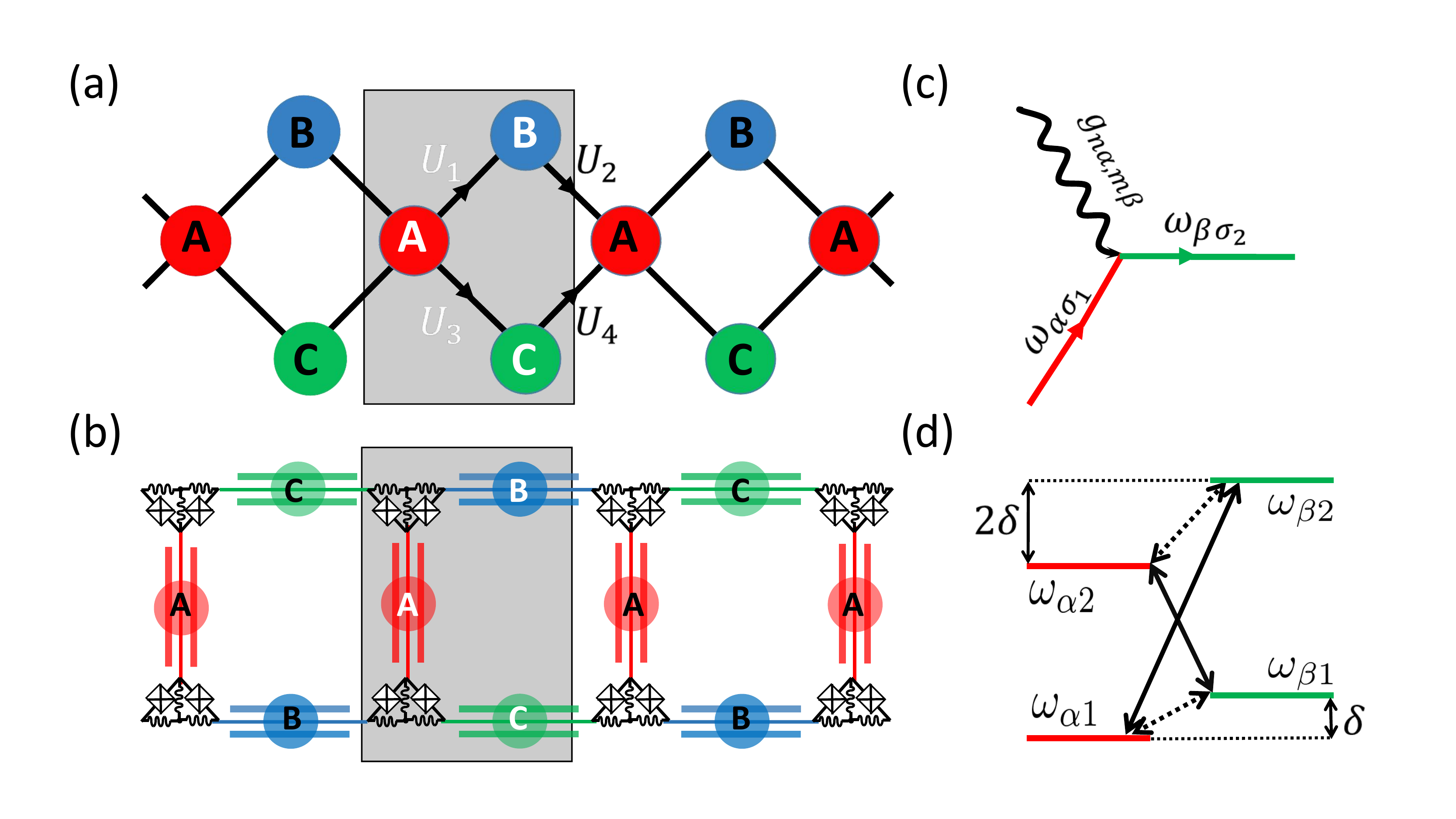}
\end{center}
\caption{(Color Online) (a) Sketch of the 1D periodic rhombic lattice composed of unit-cells with three sites labeled A, B, and C. (b) Circuit QED implementation of the proposed two component rhombic lattice. The lattice is built by the TLRs grounded at their common ends (the big dots) and connected by the coupling SQUIDs (the crossed squares). The colors of the TLRs label their different lengths and consequently different eigenfrequencies. (c) Effective $[n,\alpha,\sigma_1]\Leftrightarrow[m,\beta,\sigma_2]$ mode coupling mediated by parametric modulation of the coupling SQUID. (d) Typical eigenfrequency level diagram of the neighboring $[n,\alpha]$ and $[m,\beta]$ sites, and the corresponding modulating tone configuration of the coupling SQUID. Here $\delta=\omega_{\beta 1}-\omega_{\alpha 1}$ is the frequency difference of the $\lambda/2$ modes of the two TLRs.}
\label{Fig Lattice}
\end{figure}

 In this section, the physics of non-Abelian AB caging is illustrated in the context of a periodic 1-D rhombic lattice sketched in Fig. \ref{Fig Lattice}(a) \cite{Longhi:14,MukherjeeABCRealization2018PRL,Di2018Nonlinear}, with each site consisting of $N$ (pseudo)spin modes. The Hamiltonian of the lattice in the presence of an $U(N)$ background gauge field $\mathbf{A}$ takes the form
	\begin{equation}
	\label{Eqn RhombicHamiltonian}
	H=-J \sum_{\left\langle n\alpha, m\beta\right\rangle} \alpha_{n}^{\dagger} U_{{n}\alpha, {m}\beta} \beta_{m},
	\end{equation}
where ${J}$ is the uniform positive hopping strength between the linked sites shown in Fig. \ref{Fig Lattice}(a),  $\alpha_{n}=[\alpha_{n,1},\alpha_{n,2},\cdots,\alpha_{n,N}]^T$ with $\alpha=A,B,C$ is the multi-component annihilation operator vector of the $\alpha$th site in the $n$th unit-cell, and $U_{{n}\alpha, {m}\beta}=\exp \left[{i} \int_{[m, \beta]}^{[n, \alpha]}\mathrm{d}\textbf{x} \cdot\textbf{A}(\textbf{x})\right]$ is the translational-invariant link variable describing the unitary transformation experienced by a particle when it hops from site $[m,\beta]$ to site $[n,\alpha]$ \cite{GoldmanGauge2014RPP,KogutNAGRMP1983}.

We then define the non-Abelian AB caging by the condition that the interference matrix
	\begin{equation}
	\label{Eqn NAABCCondition}
	I=\frac{1}{2}\left(U_{2}U_{1}+U_{4}U_{3}\right),
	\end{equation}
is nilpotent, i.e.
\begin{equation}
\label{Eqn NPcondition}
\exists\quad m \in \mathbb{N},\quad\mathrm{s.t.}\quad I^{m-1}\neq0,\quad  I^m=0.
\end{equation}
Here $U_{1}$, $U_{2}$, $U_{3}$, and $U_{4}$ are the rightward link variables labeled in Fig. \ref{Fig Lattice}(a). This definition can be explained intuitively as follow: Imagine there is a particle initially populated in the $[n,A]$ site highlighted in Fig. \ref{Fig Lattice}(a). Due to the geometry of the lattice, this particle can move rightward to $[n+1,A]$ via only two paths. Along these two paths, it will gain the unitary transformations $U_{\mathrm{up}}=U_{2}U_{1}$ and $U_{\mathrm{down}}=U_{4}U_{3}$ respectively, and consequently an interference described by the interference matrix $I$ in Eq. \eqref{Eqn NAABCCondition}.

For the Abelian situation $N=1$,  AB caging happens exactly when $\pi$ magnetic flux is penetrated in each loop of the lattice. The two up and down paths shown in Fig. \ref{Fig Lattice}(a) then interfere with each other destructively and result in an vanishing $I=0$. Therefore, the particle initially in the site $[n,A]$ becomes localized as it cannot spread outward to sites further than $[n \pm 1, A]$. For the non-Abelian situation $N > 1$, the localization of the particle can still happen when $I=0$, just the same as the Abelian case \cite{RashbaCage2004PRL,RashbaCage2005PRB}. However, the matrix feature of $\mathbf{A}$ now offers possibilities of much more rich physics: A matrix-formed interference matrix $I$ can be nilpotent, with c-number $I=0$ being its $N=1$ special case. This point lies in the heart of our generalization of AB caging to the non-Abelian gauge situation.

\subsection{Two examples of nilpotent interference matrices}
To validate the proposed non-Abelian AB caging concept, we consider a special example
 \begin{equation}
 \label{Eqn Npexample1}
 \begin{aligned}
 I=& \sum_{n=1}^{N-1}\ketbra{n}{n+1} = \frac{1}{2}\left[\sum_{n=1}^{N-1}\ketbra{n}{n+1} +\ketbra{1}{N}\right.\\
 &+\left.\sum_{n=1}^{N-1}\ketbra{n}{n+1}-\ketbra{1}{N}\right].
 \end{aligned}
 \end{equation}
with nilpotent power $m=N$. Here $\left|n\right\rangle$ is the $N$-column with only its $n$th item being unity and others being zero, and a candidate decomposition of $I$ into two $U(N)$ matrices (i. e. $U_{\mathrm{up}}$ and $U_{\mathrm{down}}$) is implied. The evolution of particles on the lattice still exhibits localization feature in this case. However, the non-Abelian AB cage now has an enlarged size and a spatial location depending on the initial state of the system: Let us assume that a particle is initially prepared in the $[n,A,l]$ mode with $l\in[1,N]$. Due to the form of $I$ shown in Eq. \eqref{Eqn Npexample1}, the particle will hop into the $(l-1)$th mode when it arrives at the $[n+1,A]$ site. After it reaches the $[n+1,A]$ site, it can move rightward further. The rightmost $A$ site that the particle can approach is $[n+l-1,A]$, and its further moving towards $[n+l,A]$ is suppressed. The leftward moving of the particle can be investigated in the similar way, with $I^\dagger=(U_{\mathrm{up}}^{-1}+U_{\mathrm{down}}^{-1})/2$ being used as the interference matrix due to the reversed moving direction. The particle can thus reach as far as the $[n+l-N,A,N]$ mode. The above analysis therefore indicates that the particle will experience a spin dependent asymmetric directional breath dynamics. This is different from the Abelian case, in which a particle initial prepared in $[n,A]$ site will breath in a symmetric manner between $[n,A]$ and its four neighboring $B$ and $C$ sites \cite{Longhi:14,Di2018Nonlinear}. Such symmetry is manifestly broken in the considered non-Abelian situation.

We further study a more general situation in which $I$ has nilpotent power smaller than $N$. Without loss of generality, we assume
 \begin{equation}
 \label{Eqn Npexample2}
 \begin{aligned}
 &I = \sum_{n=1}^{m-1}\ketbra{n}{n+N-m+1}\\
 &= \frac{1}{2}\left[\sum_{n=1}^{m-1}\ketbra{n}{n+N-m+1}+\sum_{n=1}^{N-m+1}\ketbra{n}{N+1-n}\right.\\
 &+\left.\sum_{n=1}^{m-1}\ketbra{n}{n+N-m+1}-\sum_{n=1}^{N-m+1}\ketbra{n}{N+1-n}\right],
 \end{aligned}	
\end{equation}
with nilpotent power $m\in (1,N)$. Re-performing the previous analysis, we find that now the spatial configuration of the non-Abelian AB cage is not only determined by the initial mode number $l$ of the particle, but also the nilpotent power $m$ of $I$. Explicitly speaking, there are six situations, with main results summarized in Tab.\ref{Tab NAABC}. For each situation, the AB cage is characterized by its size defined by the number of $A$ sites that can be populated by the particle during its evolution, and its right and left edges, defined by the rightmost and leftmost $A$ sites that can be populated by the particle during its evolution. We should emphasize that Eqs. \eqref{Eqn Npexample1} and \eqref{Eqn Npexample2} obviously do not exhaust the possible forms of the nilpotent interference matrix $I$. Meanwhile,  these two illuminating examples have already revealed several properties of the non-Abelian AB caging which are significantly different from its Abelian counterpart: The AB cage in the non-Abelian situation becomes larger, with its size and edge location depending on both the nilpotent power of the interference matrix and the initial state of the particle. From the previous derivation, we can see that these exotic new features stem from the matrix nature of the gauge field $\mathbf{A}$ and thus have not Abelian analog.

\begin{table*}[tbhp!]
        \begin{tabular}{l|c|c|c|c|c|c}	
 		\toprule
   	Nilpotent power of $I$ &   \multicolumn{3}{c|}{$m\in(1,[{N+1}/{2}])$}& \multicolumn{3}{c}{$m\in[[{N+1}/{2}],N)$} \\ \hline
    Initial mode  &$l\in[1,m]$&$l\in(m,N-m]$ &$l\in(N-m,N]$&$l\in[1,N-m]$&$l\in(N-m,m]$ &$l\in(m,N]$\\ \hline
    Size of the cage & $l$ & $1$ & $N-l+1$ & $l$ & $N$ & $N-l+1$ \\
    Right edge & $[n+l-1,A]$ & $[n,A]$ & $[n,A]$ & $[n+l-1,A]$ & $[n+l-1,A]$ & $[n,A]$ \\
    Left edge   & $[n,A]$ & $[n,A]$ & $[n+l-N,A]$ & $[n,A]$ & $[n+l-N,A]$ & $[n+l-N,A]$\\
     \botrule		
	\end{tabular}	
	\caption{\label{Tab NAABC}Spatial configuration of the non-Abelian AB cage versus the nilpotent power $m$ of the interference matrix $I$ and the mode number $l$ initially populated by the particle. The interference matrix takes the form in Eq. \eqref{Eqn Npexample2}. For $m \in (1,[(N+1)/2])$ , the particle can move neither rightward nor leftward if $l \in (m,N-m]$. The size of the AB cage in this situation is $1$, as the same as the situation of Abelian AB caging \cite{Longhi:14}. On the other hand, if $l\leq m$ (or $l>(N-m)$), the particle can move only rightward (or leftward), and reach as far as the $[n+l-1,A,1]$ (or the $[n+l-N,A,N]$) mode. The size of the AB cage in the situation is $l$ (or $N-l+1$).  Meanwhile, for $m \in [[(N+1)/2],N)$, the particle can move rightward to reach $[n-l+1,A,1]$ mode and leftward to arrive $[n+l-N,A,N]$ mode if $l\in (N-m,m]$. The size of the AB cage in this situation is $N$.  For $l\leq(N-m)$ (or $l > m$), the particle can only move rightward (or leftward) and can reach as far as the $[n-l+1,A,1]$ (or the $[n+l-N,A,N]$) mode. In this situation, the AB cage has size $l$ (or $N-l+1$).}
\end{table*}

\subsection{A minimal $U(2)$ model}
We then turn to a $U(2)$ design which can be regarded as the minimal realization of the proposed non-Abelian AB caging. Also, this is partially due to recent advances of realizing two-component SOC in artificial systems \cite{AmoSOCPRX2015,PanJWSOCScience2016,SalernoSOC2017NJP}. The link variables can be carefully set as
	\begin{equation}
	\label{Eqn U2NAABC}
	U_{1}=U_{4}=\left[\begin{matrix}
	1&0\\0&1
	\end{matrix}\right],
    U_{2}=\left[\begin{matrix}
	0&1\\1&0
	\end{matrix}\right],U_{3}=\left[\begin{matrix}
	0&1\\-1&0
	\end{matrix}\right],
	\end{equation}
	 and such that a nilpotent $I$ with $m=2$ can be achieved:
	\begin{equation}	
    \label{Eqn U2Interference}
	 I=\frac{1}{2}(U_2U_1+U_4U_3)=\left[\begin{matrix}
	 0&1\\0&0\end{matrix}\right].
	\end{equation}
 If the particle is initially prepared in the $[n,A,1]$ mode, it can move rightward and arrive the $[n+1,A,2]$ mode, but it can reach neither the $[n+2,A]$ nor the $[n-1,A]$ sites. Meanwhile, If the particle is initially prepared in the $[n,A,2]$ mode, it can move leftward and arrive the $[n-1,A,1]$ mode, but it cannot arrive the $[n-2,A]$ or the $[n+1,A]$ sites.

 The proposed model is not the only possible $U(2)$ model exhibiting non-Abelian AB caging. However, we can prove that other possible models should take a very similar form to it upon a local basis transformation: Suppose we want  an $U(2)$ model exhibiting non-Abelian AB caging. The key point is to find a nilpotent interference matrix $I$ and its decomposition into two unitary matrices $I=(U_{\mathrm{up}}+U_{\mathrm{down}})/2$ . We recognize that, any $2 \times 2$ nilpotent matrix $I$ is equivalent to an upper triangular matrix upon a certain unitary transformation $V$, i. e.
      \begin{equation}
      \label{Eqn U2}
        	V^\dagger I V=\left[\begin{matrix}
	0&\gamma\\0&0
	\end{matrix}\right],
      \end{equation}
where $V$ corresponds to the local basis choice on the $A$ site, and $\gamma$ is determined by $I$ upon a $U(1)$ phase and can be set real-valued through the change of $V$. After figuring out the possible form of $I$, we further notice that the only possible unitary decomposition of the RHS of Eq. \eqref{Eqn U2} takes the form:
       \begin{equation}
      \label{Eqn U2deco}
        	\left[\begin{matrix}
	0&\gamma\\0&0
	\end{matrix}\right]=\frac{1}{2}(U_{\mathrm{up}}+U_{\mathrm{down}})=\frac{1}{2}\left[ \left(\begin{matrix}
	0&e^{\mathrm{i} \theta}\\ e^{\mathrm{i}\psi}&0
	\end{matrix}\right) + \left(\begin{matrix}
	0& e^{\mathrm{i}\theta}\\-e^{\mathrm{i}\psi}&0
	\end{matrix}\right)     \right],
      \end{equation}
  which further restrict the norm of $\gamma$ into $|\gamma| \leq 1 $. Then the final step of constructing the desired $U(2)$ model is to select the link variables such that $U_2U_1=U_{\mathrm{up}}$ and $U_4U_3=U_{\mathrm{down}}$. The explicit matrix forms of the link variables depends on the basis choice on the $B$ and $C$ sites. Here we can see that Eqs. \eqref{Eqn U2} and \eqref{Eqn U2deco} severely limit the choice of possible $U(2)$ models to a very small region. Our choice in the manuscript corresponds to $\gamma=1$. Other choice of models can only differ from our model upon a local basis transformation by choosing a $\gamma$ with smaller norm.

\begin{figure}[tbhp]
  \begin{center}
 		\includegraphics[width=0.48\textwidth]{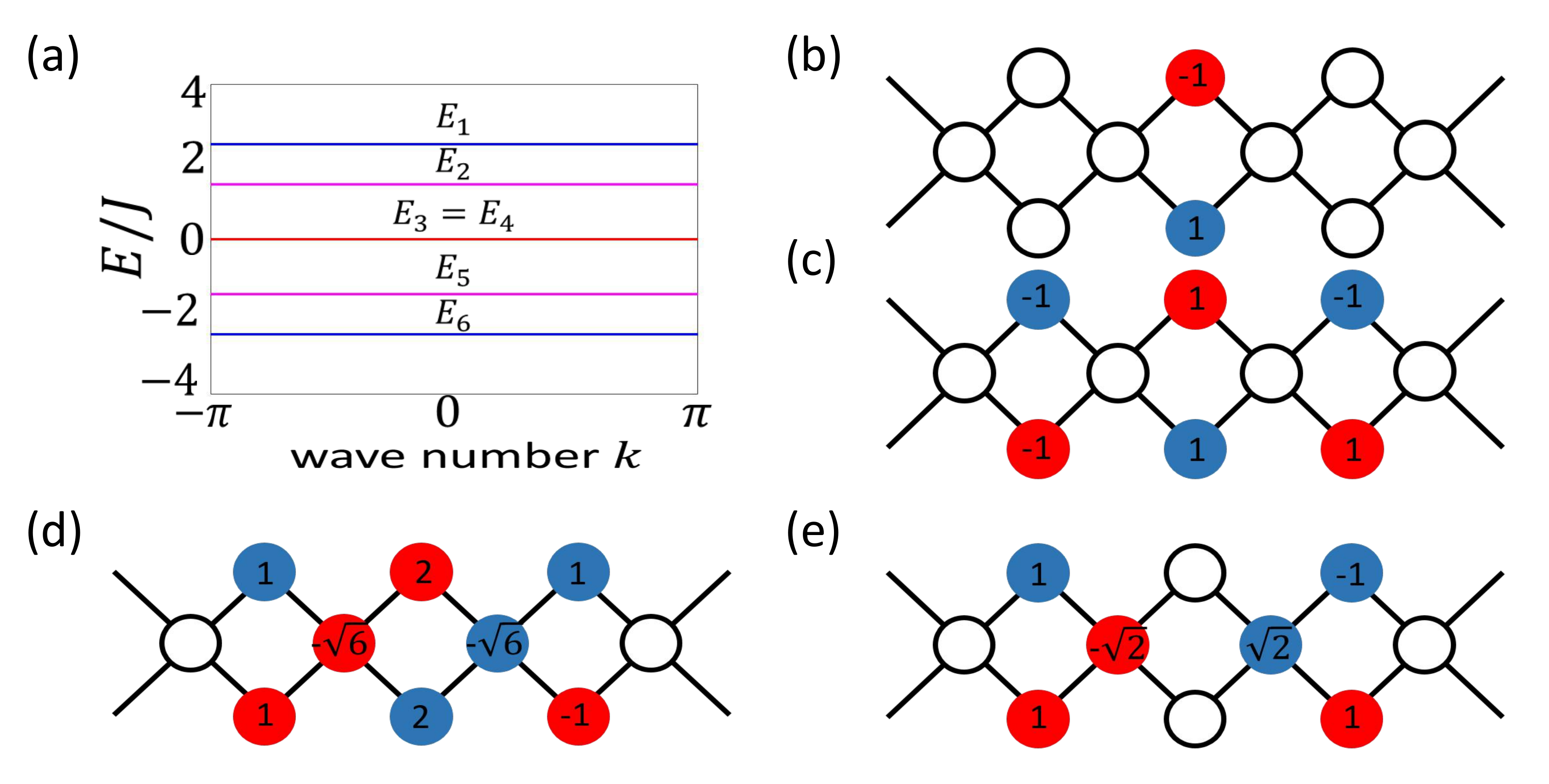}
 \end{center} 	
 	\caption{(Color Online) $U(2)$ non-Abelian AB caging in a two-component rhombic lattice with link variables defined in Eq.  \eqref{Eqn U2NAABC}. The band structure of the lattice is depicted in (a): All the six bands are completely flat, indicating localization in the absence of disorder. Four of the six un-normalized single-particle CLESs of these FBs are sketched in (b)-(e). The two-fold degenerate $E_3=E_4=0$ bands correspond to (b) and (c), while the $E_{6}=-\sqrt{6}J$, and $E_{5}=-\sqrt{2}J$ correspond to (d) and (e) respectively. The colors of the circles indicate the involved spin component (red for $\sigma=1$ and blue for $\sigma=2$), and the numbers in the circles represent the relative ratio of superposition in the corresponding CLESs.}
 	\label{Fig CLS}
 \end{figure}

 An alternative perspective is to calculate the band structure of the lattice with Eq. \ref{Eqn U2NAABC}, we can get the corresponging K-space Hamiltonian
 \begin{equation}
 H_{k}=\left[\begin{matrix}
 0&0&e^{ik}&1&1&e^{ik}
 \\0&0&1&e^{ik}&e^{ik}&1
 \\e^{-ik}&1&0&0&0&0
 \\1&e^{-ik}&0&0&0&0
 \\1&e^{-ik}&0&0&0&0
 \\e^{-ik}&1&0&0&0&0
 \end{matrix}\right],
 \end{equation}
  As shown in Fig. \ref{Fig CLS}(a), the six bands of the lattice become all completely flat, implying that a particle on the lattice should move with velocity $v=\partial E/\partial k=0$ \cite{AshcroftMermin}. We also offer a brief remark on the symmetry of the model \cite{RevModPhys.88.035005}. The flat band structure implies that the proposed $U(2)$ model does not preserve time reversal symmetry(TRS) if the two components are spin-$\frac{1}{2}$ components, because the flat bands do not exhibit Kramers degeneracy at the high symmetry points of the K-space. However, TRS is preserved if the two components are pseudospin components, because in this situation TRS corresponds to complex conjugation and all the link variables in our $U(2)$ model are real-valued. Moreover, the model has chiral symmetry(CS) where the CS operator can be described as
  \begin{equation}
   C=\left[\begin{matrix}
   \label{c}
  1&0&0&0&0&0
  \\0&1&0&0&0&0
  \\0&0&-1&0&0&0
  \\0&0&0&-1&0&0
  \\0&0&0&0&-1&0
  \\0&0&0&0&0&-1
  \end{matrix}\right].
  \end{equation}
  The existence of CS is intuitive, as Fig. \ref{Fig CLS}(a) indicates that the eigenenergies $E(k)$ and $-E(k)$ come in pair. The particle-hole symmetry(PHS) can then be defined by the combination of TRS and CS. PHS dose not existe if the two components are spin-$\frac{1}{2}$ components, and the PHS does exist with $PHS =1$ if the two components are pseudospin components. In addition, the compact localized eigenstate (CLES) for each eigenenergy is calculated and shown in Figs. \ref{Fig CLS}(b)-(e) and Figs. \ref{Fig stead_state}(a)-(b). For the two-fold degenerate middle bands with $E_3=E_4=0$, the CLES do not have $A$ site component (Figs. \ref{Fig CLS}(b) and (c)). Also, it can be checked that they are the eigenstates of the CS operator in Eq. \eqref{c}. Meanwhile, for the other four bands with eigenenergies $E_1=-E_6=\sqrt{6} J$ and $E_2=-E_5=\sqrt{2} J$, the corresponding CLES containing the $[n,A,1]$ and $[n+1,A,2]$ mode components are explicitly shown in Figs. \ref{Fig CLS}(d)-(e) and \ref{Fig stead_state}(a)-(b), respectively.

Before proceeding, we offer another discussion on the difference between our model in Eq. \eqref{Eqn U2NAABC} and the Rashba caging investigated in Refs. \cite{RashbaCage2004PRL,RashbaCage2005PRB}. As we have stated previously, flat-band localization of particles can certainly happen with vanishing interference matrix $I=0$. However, the matrix feature of $I$ can offer possibilities of much more rich physics: A non-zero interference matrix $I$ can be nilpotent, and this nilpotency leads to the flat band localization of particles. This point thus helps us to classify these two models: As mentioned in Refs. \cite{RashbaCage2004PRL,RashbaCage2005PRB}, flat-band localization can be induced in a two-component rhombic lattice by Rashba SOC with specific SOC strength. The interference matrix in that situation is completely zero, i. e. it belongs to the vanishing interference matrix $I=0$ case. On the other hand, the interference matrix calculated from Eq. \eqref{Eqn U2NAABC} is non-zero with nilpotent power $2$, i. e. it belongs to the nonzero nilpotent $I$ case.

\section{A Candidate Circuit QED Implementation}
   \label{Sec Implementation}	
\subsection{The circuit QED lattice proposal}
In this section, we consider the circuit QED lattice shown in Fig.~\ref{Fig Lattice}(b) as the candidate realization of  the proposed non-Abelian AB caging described in Eqs. \eqref{Eqn U2NAABC} and \eqref{Eqn U2Interference}. This circuit QED lattice consists of three types of superconducting transmissionline resonators (TLRs) \cite{WallraffNature2004} differed by their lengths and placed in an interlaced form. These TLRs play the corresponding roles of the A, B, and C sites of the rhombic lattice depicted in Fig.~\ref{Fig Lattice}(a). In the bulk of the TLRs, the propagation of the voltage and current fluctuations obey the wave equation \cite{YaleCQEDPRA2004}, while at their ends, the TLRs are grounded by inductors with inductances much smaller than those of the TLRs, which impose the consequent low-voltage shortcut boundary conditions for the TLRs \cite{NISTParametricConversionNP2011,FelicettiPRL2014,WangYPChiral2015,WangYPNPJQI2016,YangZHLieb2016PRA}. Therefore, the eigenmodes of the resonators are approximately their $\eta\lambda/2$ modes of the waveguide with $\eta$ being integers. For each TLR, its lowest $\lambda/2$ and $\lambda$ modes are selected as the first and the second pseudospin component, respectively. Through the parameter setting of the circuit and the external pumping pulses described in what follows, we expect that the excitation of the higher frequency modes is effectively suppressed and thus the two-mode resonator approximation is valid. The lattice can then be described by the Hamiltonian
\begin{equation}
H_\mathrm{S}= \sum_{\alpha,n,\sigma} \omega_{\alpha\sigma} \alpha_{n\sigma}^\dagger \alpha_{n\sigma},
\label{Eqn CavityHami}
\end{equation}
where the eigenfrequencies of the cavity modes are specified as
\begin{subequations}
  \label{Eqn FrequencyCondition}
  \begin{alignat}{2}
  \left[\omega_{ {A}1},\omega_{ {B}1},\omega_{ {C}1}\right]&=\left[\omega_{0}-\Delta,\omega_{{0}}, \omega_{{0}}+\Delta\right],\\
  \left[\omega_{ {A}2},\omega_{{B}2},\omega_{{C}2}\right]&=2\left[\omega_{{A}1},\omega_{ {B}1},\omega_{ {C}1}\right],
\end{alignat}
\end{subequations}
with $\omega_{{0}}/2\pi \in \left[5,6\right]\,\mathrm{GHz}$ and $\Delta/2\pi \in \left[1,2\right]\,\mathrm{GHz}$. Such configuration is for the following application of the PFC method and can be precisely realized in experiments through the length selection of the TLRs in the millimeter range \cite{NISTParametricConversionNP2011,UnderwoodDisorderPRA2012}.

We further consider the implementation of the required link variables on the lattice, taking the form in Eq. \eqref{Eqn U2NAABC} in the rotating frame of ${H}_{\mathrm{S}}$. Here we employ the dynamic modulation method \cite{NISTParametricConversionNP2011,NISTCoherentStateAPL2015,NISTHongOuMandelPRL2012,GrossPFCEPJQT2016,RoushanChiral2017NP}, which is different from the on-site modulation method used in Refs. \cite{Longhi:14, PhysRevLett.121.075502} in the sense that the a.c. modulation here is imposed on the inter-site links. This scheme has already been exploited in an experiment of generating artificial Abelian gauge field in a three-qubit ring superconducting quantum circuit \cite{RoushanChiral2017NP}. The essential physics can be illustrated intuitively through a toy model of two coupled cavities: Our aim is to implement a controllable inter-cavity photon hopping process, described by the effective Hamiltonian
\begin{equation}
{H}_{\mathrm{eff}}^{12}=Ja_{1}^{\dag}a_{2} e^{i\theta_{12}}+\mathrm{h.c.},
\label{equ:model}%
\end{equation}
where $a_{i}/a_{i}^{\dag}$ are the annihilation/creation operators of the $i$th cavity for $i=1,2$, $J$ is the effective $1\leftrightarrow 2$ hopping rates, and $\theta_{12}$ is the corresponding hopping phase. To achieve this goal, we consider a two-cavity physical Hamiltonian
\begin{equation}
{H}_{TC}={H}_{0}+{H}^{12}_{\mathrm{a.c.}}(t),\label{equ:model2}
\end{equation}
with
\begin{align}
&  {H}_{0}=\Sigma_{i=1}^{2}\omega_{i} a_{i}^{\dag}a_{i},\\
&  {H}^{12}_{\mathrm{a.c.}}(t)=g_{12}(t)[a_{1}^{\dag}+a_{1}][a_{2}^{\dag}+a_{2}],
\end{align}
where $\omega_{i}$ is the eigenfrequency of the $i$th cavity, $g_{12}(t)$ is the tunable coupling constant between the cavities, and ${H}^{12}_{\mathrm{a.c.}}(t)$ takes the similar form to the inductive current-current coupling of two TLRs which will be discussed later. Here we also assume that $g_{12}(t)$ can be tuned harmonically. This corresponds to the a.c. modulation of the coupling SQUID between the neighboring TLRs. Moreover, we assume that the parameters in ${H}_{TC}$ satisfy the far off-resonance condition:
\begin{equation}
|\omega_{2}-\omega_{1}| \gg |g_{12}(t)|.
\end{equation}

If $g_{12}(t)$ is static, the $1\leftrightarrow2$ photon hopping can hardly happen because the two cavities are far off-resonant. Meanwhile, we can implement the effective $1\leftrightarrow2$ photon hopping by modulating $g_{12}(t)$ dynamically as
\begin{equation}
g_{12}(t)=2J\cos[(\omega_{1}-\omega_{2})t-\theta_{12}]. \label{equ:driving}
\end{equation}
Physically speaking, $g_{12}(t)$ carries energy quanta filling the gap between the two cavity modes. For a photon initially placed in the 1st cavity, it can absorb an energy quantum $|\omega_{2}-\omega_{1}|$ from the $1\leftrightarrow2$ link, convert its frequency to $\omega_{2}$, and hop finally into the 2nd cavity. We can further describe this process in a more rigorous way: in the rotating frame with respect to ${H}_{0}$, ${H}^{12}_{\mathrm{a.c.}}(t)$ becomes
\begin{align}
{H}^{12}_{\mathrm{eff}}=e^{iH_{0}t}{H}^{12}_{\mathrm{a.c.}}(t)e^{-iH_{0}t}\approx g_{12} a_{1}^{\dag
}a_{2} e^{i\theta_{12}}+\mathrm{h.c.}, \label{equ:Hamiltonianchiral0}%
\end{align}
where the other $a_1^\dagger a_2^\dagger+\mathrm{h.c.}$ term are fast oscillating in the rotating frame and thus are safely neglected.

From Eq. \eqref{equ:Hamiltonianchiral0}, we notice that both the effective hopping strength and the hopping phase can be controlled by the modulating pulse $g_{12}(t)$. In particular, the control of the hopping phase $\theta_{12}$ is important as we are using this method to synthesize artificial gauge fields. Also, we notice that the only approximation we exploit in getting the effective Hamiltonian is the rotating wave approximation, which helps us to drop the off-resonant, fast oscillating terms. Due to the fact that the a.c. modulation is now exerted on the inter-site link. The Bessel function constants from the expansion of oscillating exponential functions of the form $\exp[i\cos(\omega t)]$  in the on-site modulation proposals \cite{Longhi:14, PhysRevLett.121.075502} do not appear in the effective Hamiltonian.

To facilitate the application of the described PFC method, the TLRs on the lattice are connected at their ends by connecting superconducting quantum interference devices (SQUIDs) (Fig. \ref{Fig Lattice}(b)) \cite{GrossPFC2013PRB,GrossPFCEPJQT2016}, which correspond to the linking bonds shown in Fig. \ref{Fig Lattice}(a) one-to-one. It will be figured out in the following that each of the bonds on the lattice can be independently controlled by the modulating pulse threaded in the corresponding connecting SQUID, leading to the site-resolved arbitrary control of the synthesized non-Abelian gauge field. In addition, estimations in our previous works have shown that this method can result in effective hopping strength in the range $J/2\pi \in \left[ 5, 15 \right]\, \mathrm{MHz}$, which is robust against unavoidably imperfection factors in realistic experiments including the fabrication errors of the circuit and the background low-frequency noises \cite{WangYPChiral2015,WangYPNPJQI2016,YangZHLieb2016PRA}.

\subsection{The PFC scheme of realizing non-Abelian gauge}

The essential implementation of the PFC method can be illustrated step-by-step by investigating the a.c. modulation of a particular $[n,\alpha] \Leftrightarrow [m,\beta]$ coupling SQUID. The physical coupling between these neighboring TLRs is established through the current dividing mechanism recently implemented in experiment \cite{RoushanChiral2017NP}: Consider the two TLRs located at neighboring sites $[n,\alpha]$ and $[m,\beta]$. The small grounding inductances of the two TLRs create low voltage nodes at their ends, and their
neighboring low voltage nodes are connected by the $[n,\alpha]\leftrightarrow [m,\beta]$ coupling SQUID. The connecting SQUID can be regarded as an inductance which can be tuned by the penetrating flux bias in the coupling SQUID loop at very high frequencies \cite{DCEexperimentNature2011}. The inter-TLR coupling can be understood in an intuitive way \cite{MartinisXmonPRA2015}: An excitation current from the $[n,\alpha]$ TLR, taking the form $I_{n,\alpha}\simeq \sum_{\sigma} I_{\alpha,\sigma}(\alpha_{n,\sigma}^{\dagger }+\alpha_{n,\sigma})$, will mostly flow through its grounding inductance to ground, with a small fractions $I_{n\alpha,m\beta}$ flowing to its neighboring TLR $[m,\beta]$ through the tunable $[n,\alpha]\leftrightarrow [m,\beta]$ coupling SQUID. Depending on the ratio of the Josephson inductance of the $[n,\alpha]\leftrightarrow [m,\beta]$ coupling SQUID to the grounding inductance of the $[n,\alpha]$ TLR, $I_{n\alpha,m\beta}$ will in turn flow through the grounding inductance of the $[m,\beta]$ TLR. Therefore, the $[n,\alpha]\leftrightarrow [m,\beta]$ inter-TLR coupling can be induced by such inductive current-current coupling mechanism and can be written as
\begin{equation}
\label{Eqn ACModulation}
 {H}_\mathrm{a.c.}^{n\alpha,m\beta}=g_{n\alpha,m\beta}(t)\sum_{\sigma,\sigma^\prime}(\alpha_{n\sigma}+\alpha_{n\sigma}^\dagger)(\beta_{m\sigma^\prime}+\beta_{m\sigma'}^\dagger).
\end{equation}
where $g_{n\alpha,m\beta}(t)$ is the inter-TLR coupling strength which can be a.c. modulated through the flux bias penetrated in the coupling SQUID. Here we assume that the parameters in Eq. \eqref{Eqn ACModulation} satisfy the condition
	\begin{equation}
\label{Eqn Parametercondition}
     |\omega_{\alpha\sigma}-\omega_{\beta\sigma'}| \gg |g_{n\alpha,m\beta}(t)|.
	\end{equation}
This assumption is needed for the future application of rotating wave approximation in deriving the effective Hamiltonian.

In the first step, we aim at constructing a concrete $[n,\alpha,\sigma_1]\Leftrightarrow [m,\beta,\sigma_2]$ photon hopping. This task can be mapped to the discussed toy model by setting $1\rightarrow [n,\alpha,\sigma_1]$ and $2\rightarrow [m,\beta,\sigma_2]$: As the $[n,\alpha,\sigma_{1}]$ and $[m,\beta,\sigma_{2}]$ modes are far off-resonant, the desired hopping can hardly be achieved with static $g_{n\alpha,m\beta}(t)$. Meanwhile, we can complete this task by dynamically modulating $g_{n\alpha,m\beta}(t)$ as
	\begin{equation}
	g_{n\alpha,m\beta}(t)=2J\cos[(w_{\alpha\sigma_{1}}-w_{\beta\sigma_{2}})t-\theta_{n\alpha\sigma_{1},m\beta\sigma_{2}}].
	\end{equation}
This is the process depicted in Fig. \ref{Fig Lattice}(c), where a photon initially in the $[n,\alpha,\sigma_{1}]$ mode changes its energy by absorbing/emitting an energy quanta $|w_{\beta\sigma_{2}}-w_{\alpha\sigma_{1}}|$ from/to the oscillating $g_{n\alpha,m\beta}(t)$ and then hops into the $[m,\beta,\sigma_{2}]$ mode.

More rigorously, we expand the $[n,\alpha]\leftrightarrow [m,\beta]$ coupling in Eq. \eqref{Eqn Parametercondition} in the rotating frame of $H_{\mathrm{S}}$ in Eq. \eqref{Eqn CavityHami} and get 32 terms: two from the positive and negative frequency choices of $g_{n\alpha,m\beta}(t)=J\exp[-\mathrm{i}\theta_{n\alpha\sigma_{1},m\beta\sigma_{2}}] \exp[\mathrm{i}(w_{\alpha\sigma_{1}}-w_{\beta\sigma_{2}})t]+\mathrm{h.c.}$, four hopping branches with two choices of $\sigma$ and two choices of $\sigma^\prime$, and four terms for each hopping branch,taking the form
\begin{equation*}
[\alpha_{n\sigma}\exp{-\mathrm{i}\omega_{\alpha\sigma}t}+\mathrm{h.c.}][\beta_{m\sigma^\prime}\exp{-\mathrm{i}\omega_{\beta\sigma^{\prime}}t}+\mathrm{h.c.}].
\end{equation*}
We then take a close look at the energy spectrum of the two TLRs. The typical eigenfrequency level diagram of the neighboring $[n,\alpha]$ and $[m,\beta]$ sites is shown in Fig. \ref{Fig Lattice}(d). With out loss of generality, let us set the paramter according to the Eq. \eqref{Eqn FrequencyCondition}. Therefore the frequencies of the eight hopping branches are $\Delta, 2\Delta, \omega_0+2\Delta, \omega_{0}-\Delta$, respectively. If the modulating frequency of  $g_{n\alpha,m\beta}(t)$ is resonant with one of the four hopping branches, it is off-resonant with the other three hopping branches with a detuning at least of the order $\Delta$. Therefore we look back to the expansion: the $16$ Bogoliubov terms carries frequency at least of the order $\omega_0$, they oscillate in the fastest way and should be dropped in the very first step. Then, with this observation, we can find out that only two terms are time independent, other terms carries fast oscillating factors, six with frequencies of the order $\Delta$, and eight with the frequencies of the order $\omega_0$. Since we have assumed $\omega_0 \gg \Delta \gg  J$, we dropped these fast oscillating terms and finally get the effective Hamiltonian.
	 \begin{equation}
	 \label{Eqn EffHami}	 H_{\mathrm{eff}}^{n\alpha\sigma_{1},m\beta\sigma_{2}}=J\alpha_{n\sigma_{1}}^\dagger\beta_{m\sigma_{2}}\exp(-i\theta_{n\alpha\sigma_{1},m\beta\sigma_{2}})+\mathrm{h. c.}.
	\end{equation}
 Here we can see that both the phase and the amplitude of the effective $[n,\alpha,\sigma_{1}]\leftrightarrow[m,\beta,\sigma_{2}]$ hopping can be independently controlled by the $[n,\alpha]\Leftrightarrow[m,\beta]$ modulating pulse $g_{n\alpha,m\beta}(t)$.

This parametric formalism can then be generalized to establish the matrix-form non-Abelian link variables required in Eq. \eqref{Eqn U2NAABC}. In this situation, the coupling strength $g_{n\alpha,m\beta}(t)$ in Eq. \eqref{Eqn ACModulation} contains multiple frequencies, with each tone controlling one hopping branch in the linking variable. This is schematically sketched in Fig. \ref{Fig Lattice}(d), where the red and blue lines label the eigenfrequencies of the $[n,\alpha]$ and $[m,\beta]$ sites, and the dashed and the solid arrows represent the component-preserving $U_1$ and $U_4$ and the compoent-mixing $U_2$ and $U_3$, respectively. The major obstacle of the generalization is the cross talk effect, i.e. a particular tone in $g_{n\alpha,m\beta}(t)$ may induce other unwanted hopping process which corresponds to the fast oscillating terms which are dropped in the derivation of the effective Hamiltonian. For instance, if we drive $g_{n\alpha,m\beta}(t)$ with frequency $2\Delta$ in order to establish the $[n,\alpha,2]\leftrightarrow[m,\beta,2]$ hopping, this driving can also induce the $[n,\alpha,1]\leftrightarrow[m,\beta,1]$ hopping in an off-resonant manner with detuning $\Delta$. This process is fast oscillating and dropped in the derivation of the effective Hamiltonian. However, when we calculate the Dyson series up to the second order, this process can give a second order a.c. Stark energy shift of the involved modes. Physically, we can imagine that a particle initially populated in the $[n,\alpha,1]$ mode can absorb a $2\Delta$ frequency photon from $g_{n\alpha,m\beta}(t)$ and hop virtually to $[m,\beta,1]$. However this population is not stable because it is not energy-conserving. Therefore the only fate of the particle is that it emits the absorbed $2\Delta$ photon back to  $g_{n\alpha,m\beta}(t)$ and jump back to $[n,\alpha,1]$. This is a second order perturbation process and it results in an a.c. Stark shift of the eigenenergy of the modes $\sim J^2/\Delta$. Meanwhile, this effect can be effectively suppressed as follow. As shown in Eq. \ref{Eqn U2NAABC}, the link variables have been selected to take relatively simple forms. Each link variable matrix contains at most $2$ non-zero items, implying that any coupling strength $g_{n\alpha,m\beta}$ consists of at most $2$ tones. For instance, the a.c. modulation of the $[n,A]\Leftrightarrow [n,B]$ connecting SQUID have two tones with frequencies $2\Delta$ and $\Delta$, which induce the $[n,A,1] \Leftrightarrow [n,B,1]$ and the $[n,A,2] \Leftrightarrow [n,B,2]$ PFC bonds by bridging their frequency gaps, respectively. Moreover, the eigenfrequencies of the cavity modes are set such that the two tones are significantly different. This has already been indicated in Eqs. \eqref{Eqn FrequencyCondition} and \eqref{Eqn Parametercondition} and shown in Fig. \ref{Fig Lattice}(d). In this situation, the controlling tone of one hopping branch can hardly influence the other, resulting in arbitrary control of the linking variables. With these strategies exploited, the leading effect is the a.c. Stark shifts of the cavity mode frequencies induced by the off-resonant pumping, which is of second-order and can be further compensated by the adjustment of the frequencies of $g_{n\alpha,m\beta}(t)$ which corresponds to the renormalization of $H_\mathrm{S}$ in Eq. \eqref{Eqn CavityHami}, as implied in the derivation of Eq. \eqref{Eqn EffHami}.

\subsection{Measurement of the non-Abelian AB caging}
The proposed non-Abelian AB caging physics can be observed through the coherent monochromatic pumping of a particular $[n,A,\sigma]$ mode on the lattice, which can be described by
\begin{equation}
H_{\mathrm{pump}}=P^\dagger \alpha e^{-i\Omega_{P}t}+\mathrm{h.c.},
\end{equation}
 where $\alpha$ is the vector of the annihilation operators of the whole lattice, $P$ is the corresponding pumping strength vector, and $\Omega_{P}$ is the detuning of the pumping frequency with respect to $\omega_{A\sigma}$. The pumping can inject photons into the lattice which will then experience the exotic localization dynamics analyzed in the previous section. Therefore, we expect that information about the proposed non-Abelian AB caging can be extracted from the driven-dissipation steady state of the lattice.

\begin{figure}[tbh]
	\begin{center}
		\includegraphics[width=0.48\textwidth]{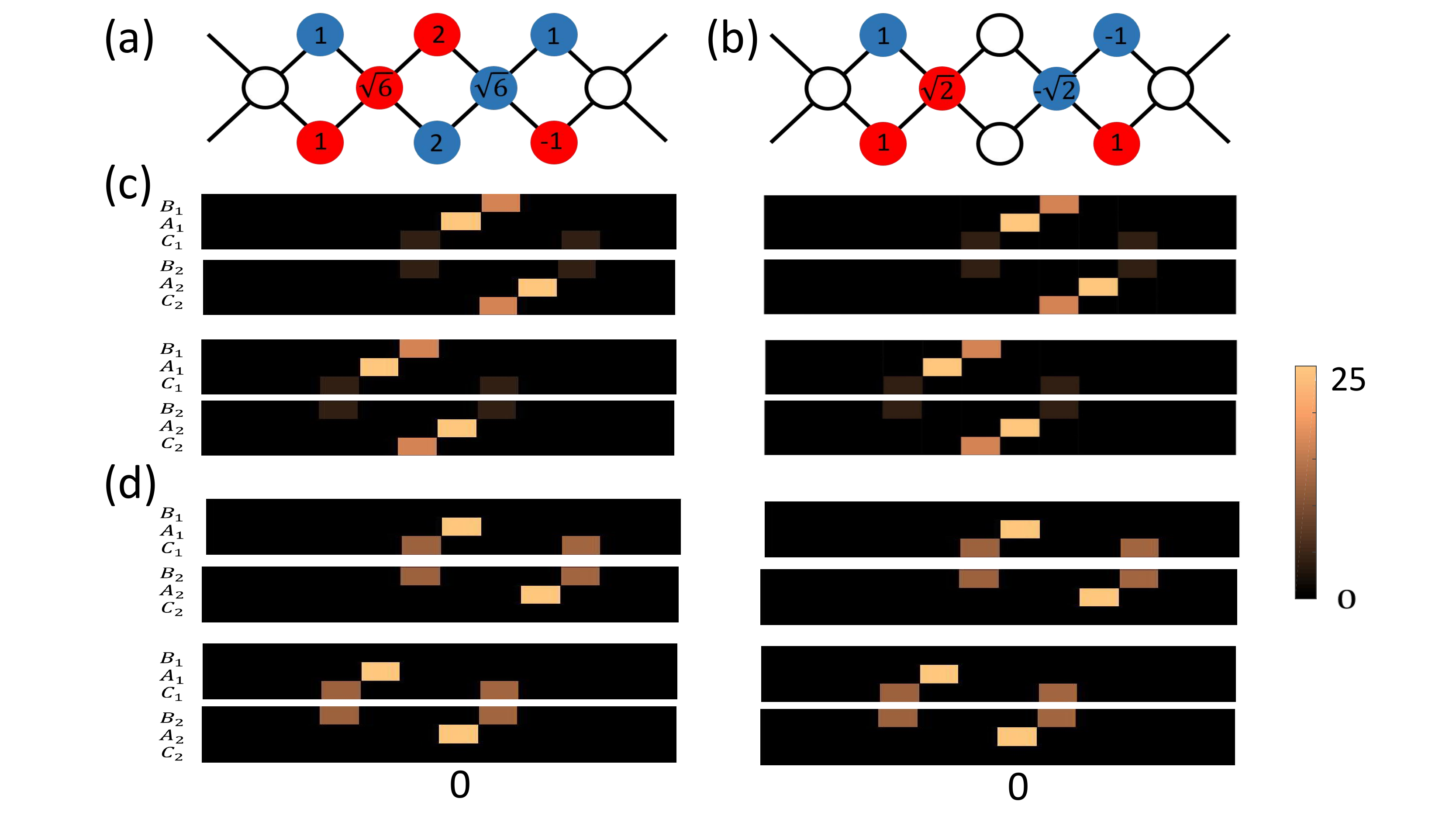}
	\end{center}	
	\caption{CLESs and the corresponding SSPN distribution of the lattice with the $[0,A]$ site being pumped. The single-particle CLESs of these FBs in (a) and (b) correspond to $E_{1}= \sqrt{6}J$ and $E_{2} = \sqrt{2}J$, respectively.  The detunings and the pumped modes are selected as: (c). $\Omega_P=\sqrt{6}J$, $[0,A,1]$/$[0,A,2]$; (d). $\Omega_P=\sqrt{2}J$, $[0,A,1]$/$[0,A,2]$. For each subfigure, the left and right panel corresponds to the SSPN calculated by using the full time-dependent physical Hamiltonian and the effective Hamiltonian.}
	\label{Fig stead_state}	
\end{figure}

 We then numerically calculate the SSPN distribution of this lattice in the presence of driving and dissipation, with results shown in Fig. \ref{Fig stead_state}(c)-(d). Roughly speaking, there are two methods of getting the SSPN. The first corresponds to the solution of the equation
 \begin{equation}
 i \frac{\mathrm{d}\langle\mathrm{\alpha}\rangle}{\mathrm{d} t}=\left[\mathcal{B}-\left(\Omega_{P}+\frac{1}{2} i \kappa\right) \mathcal{I}\right]\langle\mathrm{\alpha}\rangle+P= 0,
 \end{equation}
 where the matrix $\mathcal{B}$ is defined by $\alpha^\dagger \mathcal{B} \alpha=H$ in Eq. \eqref{Eqn RhombicHamiltonian} and $\kappa$ is the assumed uniform decay rate of the cavity modes, and the second one is solving the full time-dependent Schrodinger equation in the driven-dissipative setting for a sufficiently long time based on the physical time dependent Hamiltonian. Both method are exploited with very similar results obtained. These results thus partially validate the rotating wave approximation we have employed in getting the effective Hamiltonian in the previous section. In both simulation we consider a chain of 11 unit-cells with the pumped $A$ site at the central $x=0$ position. The parameters are chosen as $J/2\pi=|P|/2\pi=10\quad \mathrm{MHz}$ and $\kappa=0.1 J$. To maximize the injection of photons into the lattice, we set $\Omega_P$ to be resonant with one of the the eigenenergies shown in Fig. \ref{Fig CLS}(a). Explicitly, Figs. \ref{Fig stead_state}(c) correspond to $\Omega_P=\sqrt{6}J$ and pumping the$[0,A,1]$/$[0,A,2]$ modes, and Figs. \ref{Fig stead_state}(d) correspond to $\Omega_P=\sqrt{2}J$ and pumping the $[0,A,1]$/$[0,A,2]$ modes. For each subfigure, the up and lower panel denote the SSPN distribution in the first and second components of the sites, respectively. It can be clearly observed that the calculated SSPN distributions exhibit several features of the discussed non-Abelian AB caging. In particular, the injected photons become localized with asymmetric SSPN distribution with respect to the pumped $A$ site. The SSPN distribution extend more rightward(leftward) if the pumped mode is the first(second) component, as predicted in  Sec. \ref{Sec NAABC}.

 From another perspective, the calculated SSPN distributions reflect to some extent the spatial configuration of the corresponding CLESs. This can be intuitively understood because pumping a particular cavity mode with a particular eigenfrequency corresponds to exciting the corresponding CLES containing the component of the pumped cavity mode. In this sense, the pumping of the $[0,A]$ site can them be decomposed into the pumping of the superposition of the four relevant CLESs. If the pumping is resonant with one of the four CLESs and if the dissipation is sufficiently small compared with the energy gaps, only one CLES is effectively excited. For instance, the superposition of the up and lower panel in Fig. \ref{Fig stead_state}(c) coincides exactly with the CLES shown in Fig. \ref{Fig stead_state}(a). Fig. \ref{Fig stead_state}(d) also coincide with CLES in Fig. \ref{Fig stead_state}(b) in this sense. In addition, it should be noticed that the SSPN patterns shown in Figs. \ref{Fig stead_state}(c) are different from those in Figs. \ref{Fig stead_state}(d). This difference can also be attributed to the excitation of different CLESs by using different pumping frequencies.

 The calculated SSPN distribution can be experimentally detected by the following simple measurement scheme. Each of the lattice sites sufficiently involved in the SSPN calculation is capacitively connected to an external coil with input/output ports for pumping/measurement. The steady state of the lattice can be prepared by injecting microwave pulses through the input port for a sufficiently long time. During the steady-state period, energy will leak out from the coupling capacitance, which is proportional $\omega_{\alpha\sigma}\langle \alpha_{n\sigma}^{\dagger}\alpha_{n\sigma}\rangle$ with the proportional constant determined by the coupling capacitance. The target observable $\langle \alpha_{n\sigma}^{\dagger}\alpha_{n\sigma}\rangle$ can therefore be measured by simply integrating the energy flowing to the output port in a given time duration. This measurement method has already been used in experiment with both the amplitude and the phase of a coherent state of a superconducting 3D cavity were measured \cite{NISTCoherentStateAPL2015}.

\begin{figure}[tbh]
	\begin{center}
		\includegraphics[width=0.48\textwidth]{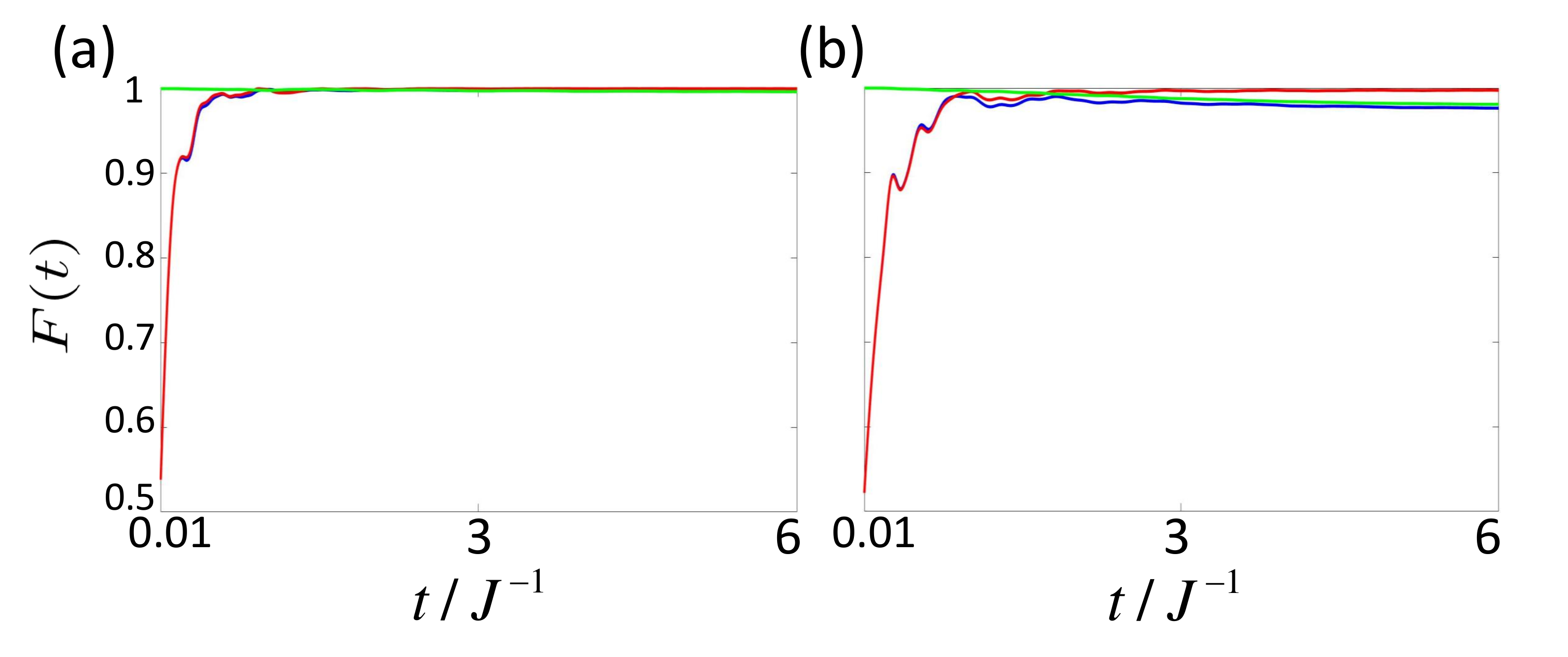}
	\end{center}	
	\caption{ Fidelity $F(t)$ with the $[0,A,1]$ mode being pumped. The detunings are selected as: (a). $\Omega_P=\sqrt{6}J$; (b). $\Omega_P=\sqrt{2}J$. In each subfigure, the time-evolving coherent state vector $\langle\alpha\rangle$ is calculated by using both the effective Hamiltonian and the full time-dependent physical Hamiltonian, and the corresponding fidelity obtained are shown by the red lines and blue lines, respectively. Moreover, the green line sketches the abstract value of normalized inner product of the two calculated coherent state vectors.}
	\label{Fig Fidelity}	
\end{figure}

 As the proposed circuit QED lattice is linear, its dynamics and steady state properties can be  well described in the framework of multi-mode coherent state. Previous theoretical and experimental papers \cite{Longhi:14, PhysRevLett.121.075502} focused mainly on the intensity property (i. e. the photon number) because the density population of particles can visualize the localization of particles directly. Meanwhile, the measurement of the phase information of the coherent state can be implemented in superconducting circuit experiments through standard microwave manipulation technique \cite{NISTCoherentStateAPL2015}. we then use a fidelity to characterize the difference between the calculated multimode coherent state of the lattice and the target CLES of the effective Hamitonian we aim to excite. This fidelity takes the form
\begin{equation}
F(t)=\frac{|\braket{a(t)}{E_n}|}{\norm{a(t)}\norm{E_n}},
\end{equation}
where $\ket{a(t)}$ is the expectation value of the annihilation operator vector of the whole lattice, and $\ket{E_n}$ is the vector describing the CLES of $H_{\mathrm{eff}}$ in the $n$th band we aim to excite. Both the time evolution of $F(t)$ under the physical time dependent Hamiltonian and the effective Hamiltonian are calculated and compared, with results shown in the Fig. 4. We see that $F(t)$ calculated from these two methods closely coincide with each other and both approach unity as $t\rightarrow \infty$. Therefore we come to the conclusion that both the amplitude and the phase information of the steady state can be effectively described by the corresponding CLES of $H_{\mathrm{eff}}$ (otherwise the calculated fidelities would deviate significantly). In addition, the coincidence of the two calculated fidelities indicates that the effective Hamiltonian is indeed correctly working.

\section{Discussion and Conclusion }

In conclusion, we have proposed in this manuscript the non-Abelian AB caging concept which is the multi-component matrix generalization of the existing Abelian AB caging concept. Distinct from its Abelian counterpart, the AB caging in this situation become sensitive to both the explicit form of the gauge potential and the initial state of the lattice. These features are the consequence of the matrix nature of our theory and thus have no Abelian analog. Moreover, we suggest a superconducting quantum circuit implementation of the proposed physics, in which the lattice sites are built by superconducting TLRs and the required non-Abelian gauge is constructed by the PFC method. The unambiguous verification of the non-Abelian AB caging can further be achieved through the steady-state manipulation of only few sites on the lattice.

While the localized steady states of the CLES excitations considered in this manuscript can be thoroughly understood in the single particle picture, what is more important is that the dispersionless flat band is an ideal platform of investigating correlated many-body states. The introduction of interaction will lead us to the realm where rich but less explored physics locates. On the other hand, as the strong coupling has already been achieved in circuit QED \cite{LiuYXReview2017PR}, the Bose-Hubbard type \cite{HuYongCrossKerr2011} and Jaynes-Cummings-Hubbard type photon-photon interaction \cite{KochReview1,KochReview2} can be incorporated by coupling the TLRs with superconducting qubits. Therefore, our further direction should be the characterization of nonequilibrium strongly-correlated photonic quantum fluids in the proposed architecture. Also, with the advances of technology, we expect the implementation of our ideas in other atomic and photonic quantum simulator platforms \cite{Yutaka}.

\begin{acknowledgments}
We thank Y. H. Wu and J. H. Gao for helpful discussions. This work was supported in part by the National Science Foundation of China (Grants No.~11774114 and No.~11874156).
\end{acknowledgments}


%

\end{document}